%% file: main.tex
\documentclass[english,aps,prper,reprint,showpacs,titlepage,longbibliography]{revtex4-2}

\usepackage[T1]{fontenc}	
\usepackage{geometry}
\usepackage{graphicx}
\usepackage{times}
\usepackage{hyperref}  
\hypersetup{colorlinks=true,urlcolor=blue,citecolor=blue,linkcolor=blue}
\usepackage{array}
\usepackage{enumerate}
\usepackage{enumitem}  
\usepackage{amsmath}
\usepackage{amssymb}
\usepackage{tikz}
\usepackage{graphicx}
\usepackage{multirow}
\usepackage{tcolorbox}
\usepackage{ragged2e}
\usepackage{tcolorbox}
\usepackage{float}
\usepackage[utf8]{inputenc}

\usepackage[normalem]{ulem}

\begin{document}
\begin{titlepage}

\title{From Self-Crafted to Engineered Prompts: Student Evaluations of AI-Generated Feedback in Introductory Physics}

\author{Amogh Sirnoorkar}

\affiliation{Department of Physics and Astronomy, \\ and  Department of Curriculum and Instruction, Purdue University, West Lafayette, Indiana - 47907.}

\author{N. Sanjay Rebello}

\affiliation{Department of Physics and Astronomy, \\ and  Department of Curriculum and Instruction, Purdue University, West Lafayette, Indiana - 47907.}


\begin{abstract}

The abilities of Generative-Artificial Intelligence (AI) to produce real-time, sophisticated responses across diverse contexts has promised a huge potential in physics education, particularly in providing customized feedback. In this study, we investigate around 1200 introductory students' preferences about AI-feedback generated from three distinct prompt types: (a) self-crafted, (b) entailing foundational prompt-engineering techniques, and (c) entailing foundational prompt-engineering techniques along with principles of effective-feedback. The results highlight an overwhelming fraction of students preferring feedback generated using structured prompts, with those entailing combined features of prompt engineering and effective feedback to be favored most. However, the popular choice also elicited stronger preferences with students either liking or disliking the feedback. Students also ranked the feedback generated using their self-crafted prompts as the least preferred choice. Students' second preferences given their first choice and implications of the results such as the need to incorporate prompt engineering in introductory courses are discussed.

\clearpage
\end{abstract}

\maketitle
\end{titlepage}

\section{Introduction}
\label{sec:intro}

Facilitating personalized student-facing feedback in large-enrollment courses can be resource-intensive in terms of both human and material resources. Consequently, several studies have explored approaches for reliably automating feedback in STEM courses~\cite{sirnoorkar2023theoretical,czajka2021novel}. Recent advancements in generative artificial intelligence (henceforth referred to as `AI')  such as ChatGPT, promise huge potential for overcoming the feedback barriers. This comes in the backdrop of  AI’s impressive ability to produce real-time, multi-modal and personalized responses based on user inputs. Accordingly, there has been a significant uptick in physics education research on AI’s ability to solve problems~\cite{kortemeyer2023could,bralin2024,polverini2024performance,zollman2024comparing,sirnoorkar2024student}, grade student responses~\cite{kortemeyer2024grading,kortemeyer2024performance,kortemeyer2025assessing,lee2024applying}, and provide customized feedback~\cite{chen2025grading,wan2024exploring}.

The process of feedback generation broadly involves an initial rubric-based grading of student responses followed by generation of reports. Research on AI grading highlights the evolving models to evaluate student responses as efficiently as, or even more accurately than humans.  For instance, Kortemeyer~\cite{kortemeyer2024grading} found GPT-4’s grading of short answers comparable to hand-engineered models, highlighting its potential for future automated grading. Chen and Wan~\cite{chen2025grading} reported GPT-4o achieving 70–80\% agreement with human graders, exceeding even inter-human agreement.  These results have also come alongside several studies reporting AI's impressive accuracy in solving introductory physics assessments~\cite{polverini2024performance,sirnoorkar2024student}.

Along the feedback front, AI has been impressive with recommendations generated from GPT-3.5 and GPT-4o in physics contexts rated either equally or better in terms of correctness and usefulness~\cite{chen2025grading,wan2024exploring}. Similar results have also been reported for AI feedback on physics tasks in languages other than English~\cite{meyer2025automatic}. The quality of generated feedback further depends on user prompts (natural language instructions to chatbots) with diligently structured  prompts yielding accurate and effective results ~\cite{meyer2025automatic,chen2025grading,wan2024exploring}. However, most of the contemporary research on AI-generated feedback in physics education remains relatively underexplored (as compared compared to research on AI's accuracy in solving physics problems or grading). Among the studies that exist, most of them have focused on empirical approaches towards designing feedback with little to no focus on incorporation of the theoretical principles of effective feedback. With students extensively employing AI to aid their learning, there is a need to explore students' approaches on seeking feedback from AI and their perspectives about AI-generated feedback from different types of prompts.

We address the above gaps in the literature by investigating student preferences about AI-feedback generated from three distinct types of prompts: (a) self-crafted prompts, (b) prompts incorporating foundational prompt-engineering techniques, and (c) prompts combining foundational techniques with principles of effective feedback. In the rest of this manuscript, we answer the following research question: 
{\em What are the broad trends in students’ preferences over AI-generated feedback based on their (a) self-crafted prompts, (b) prompts entailing foundational prompt-engineering techniques, and (c) prompts entailing foundational prompt-engineering techniques and principles of effective feedback?} 


In the next section, we provide the relevant background on prompt engineering and principles of effective feedback. In Section~\ref{sec:methods}, we describe the study’s methodology. We then present the results in Section~\ref{sec:results} followed by their discussion, implications, limitations, and future work in Section~\ref{sec:discussion-conclusion}. 

\section{Background}
\label{sec:background}

\subsection{Prompt Engineering}
\label{subsec:prompt-engineering}

Prompt engineering deals with the systematic design and optimization of prompts (the natural language instructions to chatbots) that maximize accuracy, relevance, coherence, and usability of AI-generated output~\cite{chen2025unleashing}. Broadly, prompt engineering techniques are classified into: (i) foundational and (ii) advanced. Given the study's scope, we focus only on the foundational prompting methods and recommend readers~\cite{chen2025unleashing} for advanced techniques. Below, we briefly explain the foundational techniques relevant to this study.

\begin{enumerate}[itemsep=1pt, leftmargin=*]
    \item {\em Providing instructions.} The feature of designing effective instructions to AI that yield outputs devoid of ambiguity and misinterpretations.

    \item {\em Being clear and precise.} The effective instructions to AI should be further accompanied by clear and precise commands (for instance, ``{\em Explain the meaning of resistance in the context of electric circuits.}'') as general instructions (e.g., ``{\em Explain the meaning of R in science''}) can yield contextually broad and ambiguous responses.
    
   \item {\em Role-based prompting.} Instructing AI to simulate specific roles (such as that of a domain expert) can yield task-specific and accurate outputs.  
    \item {\em Use of delimiters.} Use of symbols such as triple quotes (" " ") to separate different parts of a prompt, particularly when the prompt is complex and has multiple components. This can ensure AI to better interpret and differentiate various input elements.  

    \item {\em Zero-shot, One-shot, and Few-shot   prompting.} The technique of providing one example (`One-shot'), multiple examples (`Few-shot'), or no examples (`Zero-shot') when eliciting responses from AI platforms. When no examples are accompanied with carefully crafted prompts, AI generates responses based on its pre-trained data.
\end{enumerate}

While there are additional foundational techniques such as ``Trying several times'' and varying  the ``Temperature'' parameters, they have not been included given the current study's scope and objective. Additionally, among the three options in the fifth technique above, we employ the `Zero-shot' approach in crafting  prompts in this study. The specific prompt statement embodying the discussed foundational techniques can be found in Table~\ref{tab:prompts} (prompt for Feedback B). 

\renewcommand{\arraystretch}{1.2}
 \begin{table}[tb]
\begin{ruledtabular}
\caption{\label{tab:prompts} The three types of feedback (first column) and corresponding prompt statements (second column) embedding the foundational prompt engineering techniques and principles of effective feedback discussed in Section~\ref{sec:background}.}

\begin{tabular}{p{0.18\linewidth} p{0.8\linewidth}}

Feedback & Prompt \\
\hline

Feedback A & [Student generated prompt]. \\ 

Feedback B & """You are an expert physicist and your objective is to give feedback on my answer which is presented as an argument with a claim, evidence and reasoning about a physics problem.""" 

""The problem involves identifying the physical significance of a graph’s slope with the gravitational potential energy of a satellite (U) plotted on y-axis and the satellite’s distance from the planet’s center (r) on x-axis."""

"""The following is my answer (argument): [INSERT YOUR ARGUMENT HERE]."""

"""Provide relevant and useful feedback for my argument.""" \\

Feedback C & [Prompt for Feedback B (except for the last sentence)] + 

"""Provide feedback by explicitly highlighting the correct answer, my provided answer (including strengths and limitations), and gaps (if any) between them within 200 words. Also suggest three potential ways through a bulleted list which can help me improve my performance on similar questions in future."""
\end{tabular}
\end{ruledtabular}
\end{table}

\subsection{Principles of effective feedback}
\label{subsec:effective-feedback}

Feedback is the information provided by an agent aimed at improving specific  knowledge and/or skills contextualized in a given task~\cite{hattie2007power}. Any effective feedback communicates the following three aspects clearly~\cite{hattie2007power,sadler1989formative,black2009developing,noroozi2023gender,patchan2016nature}: 

\begin{enumerate}[leftmargin=*,noitemsep]\vspace{-.7\baselineskip}
    \item {\em Desired performance} (``{\em Where should the learner be?}'') highlighting expected or ideal response to the task. For instance, in the context of physics problem solving, this feature would correspond to the correct answer.

    \item {\em Current performance} (``{\em Where is the learner now?}'') highlighting details of the learner's current state or performance. This is often accompanied with affective features such as encouragements and positive reinforcements about the current performance that motivate learners to engage constructively with the feedback.

    \item {\em Concrete approaches to fill the gaps} (``{\em How to get there}'') between the expected and current performances. This includes cognitive features such as task descriptions,  specific areas for improvement, actionable suggestions, and plans for implementing the changes (wherever possible).
\end{enumerate}

In addition to these features, timeliness and personalization contribute to the effectiveness of feedback~\cite{carless2011developing,pardo2019using}. While real-time feedback facilitates learners to address the issues as they actively engage with the task, personally tailored feedback can contribute to its relevance and uptake. 

In this study, one of the prompts provided to students to elicit feedback from AI incorporated the above-mentioned features (prompt statement for Feedback C in Table~\ref{tab:prompts}). The timeliness and personalization aspects are addressed by AI’s ability to generate real-time and user-centric responses based on the prompt.

\section{Methods}
\label{sec:methods}

\begin{figure}[tb]
    \centering
    \begin{tcolorbox}
     \justify{You are an intern in an astrophysics laboratory, monitoring the motion of a satellite of mass 500$kg$ around a planet (with mass $6.42 \times 10^{23}$ $kg$) in a nearby galaxy. You receive data the following from the observatory highlighting the satellite’s gravitational potential energy with respect to its distance from the planet’s center.}

\noindent
\begin{tabular}{p{0.3\linewidth} p{0.28\linewidth} p{0.4\linewidth}}
\hline
  {Altitude `$h$' (km) } & Distance `$r$' (m) & Potential energy `U' (J) \\
\hline 
  100 & $3.49 \times 10^6$ & $ -6.11 \times 10^6$ \\
  200 & $3.59 \times 10^6$ & $ -5.94 \times 10^6$ \\
  300 & $3.69 \times 10^6$ & $ -5.78 \times 10^6$ \\
  400 & $3.79 \times 10^6$ & $ -5.62 \times 10^6$ \\
  500 & $3.89 \times 10^6$ & $ -5.48 \times 10^6$ \\
\end{tabular}

\vspace{-1em}
\justify{Consider a plot of $U$ ($y-$axis) vs $r$ ($x-$axis). What would be the physical significance of the graph’s slope? Detail your argument by explicitly highlighting your Claim, Evidence, and Reasoning using relevant equations, diagrams, and physics principles.}

\end{tcolorbox}
\caption{Statement of the problem for which students constructed argument and sought feedback from AI.}
\label{fig:problem-statement}
\end{figure}

Our data comes from student responses to an extra-credit activity from a large-enrollment introductory physics course at a midwestern R1 university. The course follows `Matter and Interactions' textbook from Chabay and Shwerwood~\cite{chabay2015matter} with focus on energy, momentum, and angular momentum principles. The course also entails students constructing arguments with the ``Claim, Evidence, and Reasoning (CER)''~\cite{toulmin2003uses,mcneill2008inquiry} structure as part of their assignments. The extra-credit activity was administered through Qualtrics~\footnote{https://www.qualtrics.com/} at the end of the Spring 2025 semester. The activity involved students solving a physics problem entailing hypothetical data about variation of a satellite's gravitational potential energy ($U$) with respect to its distance from a planet’s center ($r$) ( Figure~\ref{fig:problem-statement}). Given the information, students were first asked to determine the planet's radius and magnitude of acceleration due to gravity on its surface. They then constructed and submitted an argument with CER structure explaining the physical significance of the slope of the $U$ versus $r$ graph.

Students were then asked to use any AI platform of their choice for seeking feedback to their arguments in the next three tasks with the first task involving them to craft their own prompts (henceforth referred as `Feedback A'). Students then used a provided prompt template incorporating foundational prompt-engineering techniques (described in Section~\ref{subsec:prompt-engineering}) to seek feedback (`Feedback B') by inserting their arguments in the highlighted area. In the third one (`Feedback C'), students repeated the same process of the previous task, but this time with additional prompt text embodying principles of effective feedback (described in Section~\ref{subsec:effective-feedback}). Specific wording of the provided prompts in second and third tasks can be found in Table~\ref{tab:prompts}. All three feedback statements and their corresponding prompts were collected as data. Students then ranked the three feedback statements based on their usefulness and provided relevant justifications. In the end, students listed three things that they learned (if any) from this activity and the extent to which they found the entire exercise useful for their learning. In the remaining part of this paper, we focus on the data corresponding to  students' ranking of the three feedback statements and briefly refer to the justifications. 

Of the 2044 students enrolled for this course in Spring 2025 semester, 1235 had responded (60\% response rate) by the time this manuscript was being written. The responses were downloaded from Qualtrics and were tabulated in a spreadsheet. Basic spreadsheet functions (e.g., ``COUNTIF'') were used to determine the trends in students' ranking of the feedback types. These are detailed in the next section. 

\section{Results}
\label{sec:results}

\begin{figure}
    \centering
    \includegraphics[width=\linewidth]{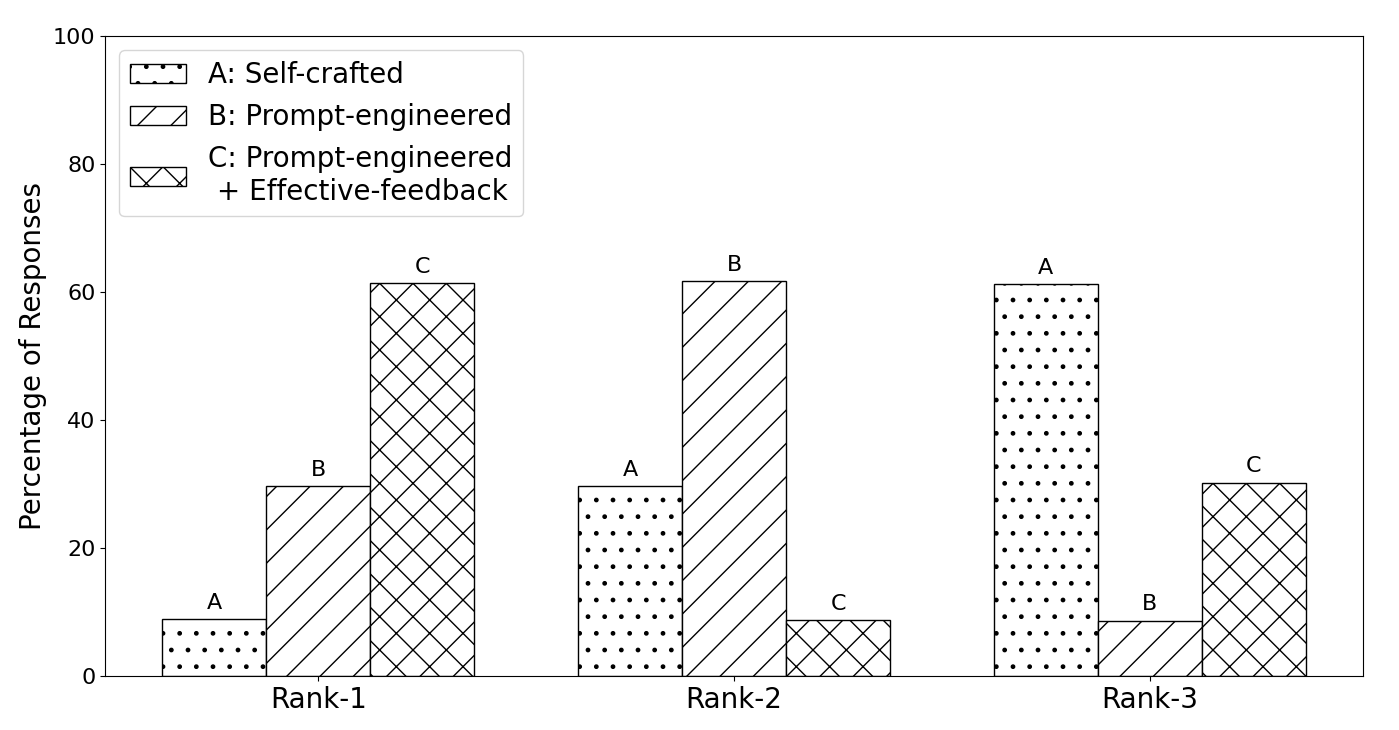}
    \caption{Plot highlighting students' ranking of the three types of feedback generated from prompts that: (i) were self-crafted, (ii) entailed features of prompt engineering, and (iii) entailed features of prompt engineering along with principles of effective feedback.}
    \label{fig:ranking}
\end{figure}

The objective of this study is to investigate students’ preferences for AI-generated feedback obtained from their (a) self-crafted prompts, (b) provided prompts entailing prompt-engineering techniques, and (c) provided prompts that entail these techniques along with principles of effective feedback. Below, we briefly discuss the overarching preferences and provide representative quotes highlighting reasoning behind the students' choice for each distinct feedback type.

Figure~\ref{fig:ranking} represents the broad trends in students' preferences across the three feedback versions. Overall, students preferred Feedback C (generated using the provided prompt having the features of both prompt engineering and effective feedback) as their top choice and rated Feedback A (generated using self-crafted prompts) as their least preferred one. Feedback B (generated from prompts entailing features of prompt engineering) tends to be the predominant second choice among the three versions. In summary, an overwhelming majority of students preferred feedback generated from structured prompts (combined preferences across Feedback types B and C). These trends are detailed below.

Across 1235 responses, around 61.4\%  of students ranked Feedback C as the most useful, followed by B (29.7\%) and A (8.9\%). For their second choice, 61.7\% chose Feedback B, followed by A (29.6\%), and  C (8.7\%). Finally as their least preferred feedback, a major section of students ranked Feedback A (61.2\%), then C (30.2\%), and lastly B (8.6\%). These results highlight three overarching trends about student preferences. Firstly, a substantial section (91.1\%) of students preferred feedback generated using structured prompts (combined first choices across B and C). Within this preference, a majority of students preferred the feedback generated using the prompt with combined features of prompt engineering and effective feedback (C). Secondly, this top choice was either liked or disliked by students. This is evident from 61.7\% indicating C as their top choice and 30.1\% preferring it as their least one. Only 8.2\% however highlighted C as their second choice. Lastly, students' least preferred feedback corresponded to the one generated using their self-crafted prompts.    

\begin{figure}
    \centering
    \includegraphics[width=\linewidth]{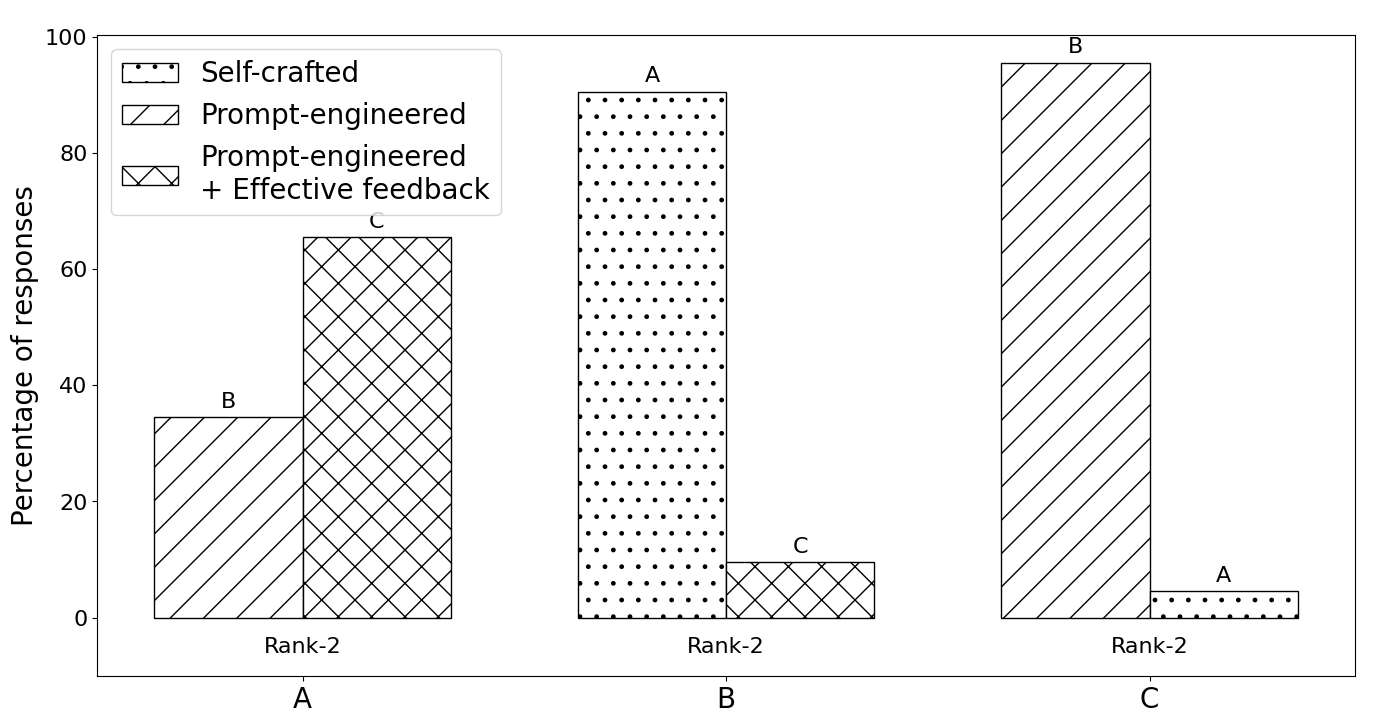}
    \caption{Plot highlighting students' second choices given their first choice of each feedback type. The feedback types at the center of the two columns on the x-axis represent the preferred feedback choice whereas those on top of the columns indicate the second choice.}
    \label{fig:rank-distribution}
\end{figure}

Beyond these overarching trends, we also investigated students’ patterns in their second choice (rank) given their first-choice of feedback type. Figure~\ref{fig:rank-distribution} illustrates these trends. Below, we briefly discuss them along with providing students' representative quotes. 

Among students who ranked  A as their top choice, two-thirds chose C as their next preferred version with remaining one third choosing B. 
Below is a quote from one of the students explaining their choice for A followed by C:

\begin{quote}
  ``{\em I think from all 3, the one I can more easily understand is feedback A, it is immediately more mathematical and if I had done something wrong it would be good to have quick understanding of where I was mistaken. But the one who highlighted more the points I need to reinforce the thesis is feedback C, as it quickly tells me I need to reinforce some points as making clear the direction of the force and the graph behavior, parts lacking in my original explanation. }''   
\end{quote}

An overwhelming 90\% of students who ranked Feedback B as their top choice, selected A, as their second preference. The remaining 10\% however chose C. 
As one of the responses following this broad trend noted:

\begin{quote}
    ``{\em A and B I found to be the most useful. I liked that A went into a lot of detail on what I did right mathematically so I could confirm that I did the problems correctly. However, B gave me the most detail on what I could improve on overall and for this problem so I ranked it as my first choice. C was too vague and not as helpful. }''
\end{quote}

Finally, 95.5\% of students who preferred Feedback C, selected B as their second choice followed by mere 4.5\% for A. 
A student who aligned with this trend mentioned:

\begin{quote}
``{\em I found C the most useful. It gave a detailed explanation for what mistake I made and gave useful answers for how I can improve. It was better than B since it contained mostly the same information, but was much more concise. Both B and C were far more detailed and more useful than A, but C was the best due to its efficient wording.}''
\end{quote}

In summary, there was no common preference in students’ second-choice selections. Those who chose A first tended to pick C second; students who ranked B highest typically selected A next; and among those who rated C as most useful, B was the most common second choice.

\section{Discussion and Conclusion}
\label{sec:discussion-conclusion}

This study investigated students' ranking of three types of feedback generated from as many distinct types of prompts. The first prompt type corresponded to students' self-crafted ones. While the second prompt type entailed the foundational techniques of prompt engineering, the third reflected a combination of these techniques with features of effective feedback. 

Results highlight that an overwhelming fraction of students preferred feedback generated using structured prompts, with those entailing combined features of prompt engineering and effective feedback to be favored most. The popular choice also elicited stronger preferences in terms of either liking or disliking. Lastly, feedback generated via self-crafted prompts was least preferred. Figure~\ref{fig:ranking} summarizes these results. Furthermore, examination of second-choice selections given each first choice shows that there was no preferred second choice. We particularly observed strong characteristic preferences for second choice when students selected B or C as their top choice. Figure~\ref{fig:rank-distribution} highlights these trends. 

Every result reported in this study uniquely contributes to the current understanding of AI-generated feedback in physics education as the research landscape pertaining to this domain remains relatively underexplored. These results also align with observations from other fields. For example, Zhang {\em et al.}~\cite{zhang2024students} examined students’ perspectives on AI feedback in programming contexts, finding a preference for feedback versions embodying theoretical principles of effective feedback. They further reported that students’ preferences were influenced by prompt characteristics and qualitative features of the feedback. Our results in the context of physics education too align with these observations.

There are several implications of the results reported in this study. For educators, students' clear preference for feedback generated by structured prompts over self-crafted ones suggests the need for effective integration of AI literacy (particularly prompt engineering) in introductory courses. This integrated approach can empower students to effectively leverage emerging AI tools for soliciting better feedback and enhancing their learning. Given students’ emphasis on feedback generated using prompts entailing effective-feedback features, the result also calls for focus on leveraging rich literature on effective feedback in designing and delivering AI feedback. 

However, the results reported in this study include several limitations. One of the underlying assumptions made in this study is that students' prompts were devoid of prompt engineering and effective feedback features, and were thus distinct from the other two prompt types. Though there is evidence in the literature in support for making this assumption~\cite{sawalha2024analyzing}, the results are still conditional on this factor. Second, we did not collect data on the specific AI model used by students for seeking feedback, and we acknowledge that the output quality varies by model. Although we suggested a particular AI platform in the prompt, students could choose any platform, and we did not record model details. This design choice represents an inherent limitation of our study.

Future work would focus on characterizing students' prompts along with their reflections about feedback preferences using systematic qualitative approaches. Such a work would shed insights on students' prompting practices as well as preferred features in each of the feedback versions. We would also seek to explore the interrelationship (if any) between the correctness of students' arguments vis-\`{a}-vis their preferences about a feedback version.

\section{Acknowledgments}

Thanks to Winter Allen, Amir Bralin, and Ravishankar Chatta Subramanium for their valuable inputs. This research is supported by Purdue University's Innovation Hub. ChatGPT-4o was used to enhance grammatical correctness.

\clearpage

\input{main.bbl}
\end{document}

%% file: main.bbl
%